\documentclass[prd,nofootinbib,preprint,superscriptaddress]{revtex4}
\usepackage[T1]{fontenc}
\usepackage[latin9]{inputenc}
\usepackage{amsmath}
\usepackage{amsthm}
\usepackage{amssymb}
\usepackage{cancel}
\usepackage{graphicx}
\usepackage[unicode=true, bookmarks=false, breaklinks=false,pdfborder={0 0 1},backref=section,colorlinks=false] {hyperref}
\usepackage{fancyhdr}
\usepackage{epsfig}
\usepackage{slashed}
\usepackage{mathrsfs}
\usepackage[thinc]{esdiff}
\usepackage[titletoc]{appendix}
\usepackage{tikzsymbols}
\usepackage{float}
\usepackage{feynmf}
\usepackage{tikzsymbols}
\usepackage{dsfont}
\usepackage{xcolor}
\usepackage{slashed}
\usepackage{braket}
\usepackage{mathrsfs}
\usepackage{xcolor}
\usepackage{gensymb}
\usepackage{latexsym}
\usepackage[compat=1.1.0]{tikz-feynman}
\begin{document}
\title{Pseudoscalar Higgs Production at Muon Colliders: The Role of One-Loop Effective Vertices}

\author{Fayez Abu-Ajamieh \footnote{Currently at  SOKN Engineering, Whitefish, Montana, USA.}}
\email{fayezabuajamieh@gmail.com, }
\affiliation{Centre for High Energy Physics, Indian Institute of Science, C V Raman Avenue, Bengaluru 560 097, India  }%

\author{Sagar Modak}%
 \email{sagarmodak@iisc.ac.in}
\affiliation{Centre for High Energy Physics, Indian Institute of Science, C V Raman Avenue, Bengaluru 560 097, India }%

\author{Samadrita Mukherjee}
\email{samadritamukherjee657@gmail.com}
\affiliation{Centre for High Energy Physics, Indian Institute of Science, C V Raman Avenue, Bengaluru 560 097, India }%

\author{Sudhir K Vempati}
\email{vempati@iisc.ac.in}
\affiliation{Centre for High Energy Physics, Indian Institute of Science, C V Raman Avenue, Bengaluru 560 097, India }%


\date{\today}

\begin{abstract} 
We investigate the production of the pseudoscalar Higgs boson $A$ at muon colliders within the framework of Type-II and Type-X Two-Higgs-Doublet Model (2HDM) at the Next-to-Leading Order (NLO), utilizing an Effective Field Theory (EFT) approach. In particular, we analyze the level of enhancement to the cross section due to the inclusion of the one-loop corrections involving $\gamma$ and $Z$ boson fusion compared to the tree level contribution. We find that for Type-II, including the effective vertices of $\gamma\gamma A$, $\gamma Z A$ and $ZZ A$, could lead to an enhancement of a factor of $\sim 2$ at low $m_A$ and low $\tan\beta$, whereas for Type-X, the enhancement could reach $\mathcal{O}(10)$ in the same regime. We also investigate the impact of the COM energy and $\tan \beta$ on the production cross section. We find that for the region of the parameter space not excluded by experiment, cross sections of $\gtrsim 1$ fb for Type-II, and $\gtrsim 5$ fb for Type-X, are possible, making the proposed muon collider a feasible alternative for probing the 2HDM extended Higgs sector.
\end{abstract}

\maketitle


\section{\label{sec:intro} Introduction}
With the discovery of the Higgs boson in the LHC \cite{ATLAS:2012yve, CMS:2012qbp}, the final piece of the SM was put in place, and the SM has been completed. Nonetheless, the SM is incapable of answering many questions that are still open, such as the masses of neutrinos, the origin of matter-antimatter asymmetry, the strong CP problem, and the nature of dark matter and dark energy to name a few. Answering these questions would require extending the SM to include new sectors, and many of these extensions require the existence of a richer Electroweak (EW) sector (see for example \cite{Abu-Ajamieh:2020yqi, Abu-Ajamieh:2021vnh, Abu-Ajamieh:2021egq, Abu-Ajamieh:2022ppp, Dawson:2022zbb, Abu-Ajamieh:2022nmt, Abu-Ajamieh:2024rzw, Abu-Ajamieh:2025vxw}).

One of the most popular extensions to the EW sector is the two-Higgs-Doublet Model (2HDM), where the SM EW sector is extended to include an additional Higgs doublet (see \cite{Branco:2011iw} for a comprehensive review). The two Higgs doublets $H_{1}$ and $H_{2}$ have hypercharges $Y=+1$ and $-1$ respectively,  
\begin{equation}\label{eq:2HD_1}
    H_{1} = \begin{pmatrix}
        \phi_{1}^{0*} \\
        -\phi_{1}^{-}
    \end{pmatrix}; 
    \hspace{5mm}    H_{2} = \begin{pmatrix}
        \phi_{2}^{+} \\
        \phi_{2}^{0}
    \end{pmatrix}.
\end{equation}

Minimizing the scalar potential leads the doublets to acquire VEVs $v_{1}$ and $v_{2}$, such that $\sqrt{v_{1}^{2}+v_{2}^{2}} = v_{\text{SM}} = 246$ GeV, and $\tan{\beta} \equiv v_2/v_1$. Expanding the fields around the VEV, the complex doublets can be expressed as
\begin{eqnarray}\label{eq:2HD_2}
  H_{1} = \begin{pmatrix}
   (v_{1} + \rho_{1} + i\eta_{1})/\sqrt{2} \\
   \phi_{1}^{-}\\
        \end{pmatrix};        
&
    H_{2} = \begin{pmatrix}
        \phi_{2}^{+}\\
        (v_{2} + \rho_{2} + i\eta_{2})/\sqrt{2}
    \end{pmatrix}; 
\end{eqnarray}
where $\rho_{a}$ are the real parts, $\eta_{a}~a = 1,2$ are the neutral pseudo-Nambu-Goldstone bosons. We can see that the 2HDM has a total of 8 degrees of freedom, 3 of which are eaten by the $W^{\pm}$ and $Z$ bosons after EW Symmetry Breaking (EWSB) becoming their longitudinal modes, leaving 5 degrees of freedom that correspond to 5 particles: Two neutral scalars, $h$ which is identified with the SM Higgs, and $H$; two charged scalars $H^{\pm}$, and one pseudoscalar particle $A$. 

A major issue with the original 2HDM is the ubiquitous Flavor-Changing Neutral Currents (FCNC), which is avoided by imposing a $Z_{2}$ symmetry on the Higgs fields, such that Yukawa interactions are restricted. The different possibilities of assigning this symmetry lead to 4 different types of the 2HDM, which are dubbed Types I, II, X and Y. These types differ by their Yukawa couplings as shown in Table \ref{tab:Yukawa_2HDM}. The extended Higgs sector of these different models have been searched for thoroughly in the LHC with no success so far (see \cite{Kling:2020hmi} and the references therein). Independently, research on 2HDM searches at future colliders has taken place; including $e^{+}e^{-}$ colliders such as the International Linear Collider (ILC) \cite{ILC:2013jhg, Behnke:2013lya, Hashemi:2018kct} and the Compact Linear Collider (CLIC) \cite{CLIC:2016zwp, Aicheler:2012bya, CLIC:2018fvx, Hashemi:2023tej}; hadron colliders such as the 100 TeV Future Circular Collider (FCC) \cite{FCC:2018vvp, FCC:2018byv} and the Super Proton Proton Collider (SPPC) \cite{Kling:2018xud, Li:2020hao,Hajer:2015gka, Craig:2016ygr}; and $e\gamma$ colliders \cite{Sasaki:2017fvk}.

\begin{table*}[!b]
\begin{center}
\begin{tabular}{ |c|c|c|c|c|c|c|c|c|c| } 
 \hline
\textbf{Type} & \textbf{$R^{A}_{u}$} & \textbf{$R^{A}_{d}$} & \textbf{$R^{A}_{l}$} &  \textbf{$R^{H}_{u}$} & \textbf{$R^{H}_{d}$} & \textbf{$R^{H}_{l}$} &  \textbf{$R^{h}_{u}$} & \textbf{$R^{h}_{d}$} & \textbf{$R^{h}_{l}$} \\
 \hline
 I & $\cot{\beta}$ & $-\cot{\beta}$ & $ - \cot{\beta}$ & $\frac{\sin{\alpha}}{\sin{\beta}}$ & $\frac{\sin{\alpha}}{\sin{\beta}}$ & $\frac{\sin{\alpha}}{\sin{\beta}}$ & $\frac{\cos{\alpha}}{\sin{\beta}}$ & $\frac{\cos{\alpha}}{\sin{\beta}}$ & $\frac{\cos{\alpha}}{\sin{\beta}}$\\
 II & $\cot{\beta}$ & $\tan{\beta}$ & $  \tan{\beta}$ & $\frac{\sin{\alpha}}{\sin{\beta}}$ & $\frac{\cos{\alpha}}{\cos{\beta}}$ & $\frac{\cos{\alpha}}{\cos{\beta}}$ & $\frac{\cos{\alpha}}{\sin{\beta}}$ &  $- \frac{\sin{\alpha}}{\cos{\beta}}$&  $-\frac{\sin{\alpha}}{\cos{\beta}}$ \\
 X & $\cot{\beta}$ & $-\cot{\beta}$ & $ \tan{\beta}$ & $\frac{\sin{\alpha}}{\sin{\beta}}$ & $\frac{\sin{\alpha}}{\sin{\beta}}$ & $\frac{\cos{\alpha}}{\cos{\beta}}$ & $\frac{\cos{\alpha}}{\sin{\beta}}$ & $\frac{\cos{\alpha}}{\sin{\beta}}$ &  $-\frac{\sin{\alpha}}{\cos{\beta}}$\\
 Y & $\cot{\beta}$ & $\tan{\beta}$ & $ - \cot{\beta}$ & $\frac{\sin{\alpha}}{\sin{\beta}}$ & $\frac{\cos{\alpha}}{\cos{\beta}}$ & $\frac{\sin{\alpha}}{\sin{\beta}}$ & $\frac{\cos{\alpha}}{\sin{\beta}}$ & $-\frac{\sin{\alpha}}{\cos{\beta}}$ & $\frac{\cos{\alpha}}{\sin{\beta}}$   \\
  \hline
\end{tabular}
\caption{The ratios of the Yukawa couplings relative to the SM Higgs Yukawa couplings in the 4 different types of the 2HDM. $\alpha$ is the neutral Higgses mixing angle and $\tan{\beta} = \frac{v_{2}}{v_{1}}$.} 
\end{center}
\label{tab:Yukawa_2HDM}
\end{table*}

Recently, there have been proposals put forward advocating the use of muon colliders operating at energies between 3 and 30 TeV \cite{Delahaye:2019omf, MuonCollider:2022xlm, Aime:2022flm, Schulte:2021hgo, Bartosik:2020xwr, Long:2020wfp, Accettura:2023ked,Han:2020uid, InternationalMuonCollider:2025sys}. Testing the 2HDM via muon colliders was suggested long ago \cite{Krawczyk:1997be}, however, recently there has been increased interest in investigating the phenomenology of the 2HDM in muon colliders. It was shown in \cite{Han:2021udl} that a muon collider has an excellent capability of discriminating the four types of the 2HDM in muon colliders for $\tan{\beta} \gtrsim 5$ via the production of heavy BSM Higgses. The phenomenology of the associated charged Higgs production with a fermion and a boson pair in muon and $e^{+}e^{-}$ colliders was studied in \cite{Ouazghour:2024twx}, where it was shown that at a muon collider with energy $\sqrt{s} = 3$ TeV, the cross section of the associated single Higgs production with a boson pair could exceed 0.1 fb. NLO EW corrections for multiple massive boson production processes at future muon colliders have been computed, including inclusive cross sections and differential distributions for $HZ$ production at various center-of-mass energies in \cite{Bredt:2022dmm}. In that study, the authors highlighted the significance of EW corrections at high energies and boson multiplicities, emphasizing the potential of muon colliders for probing the EW sector.

\begin{figure}[!t]
    \centering
\includegraphics{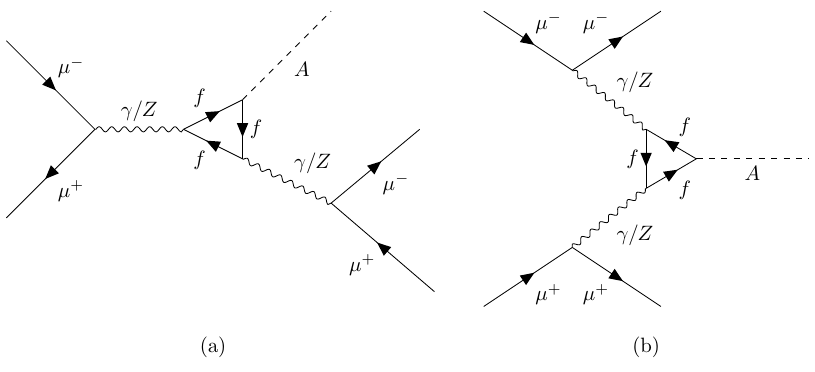}
    \caption{The production of $A$ via the fusion of $\gamma\gamma$, $ZZ$, and $\gamma Z$ through the $s$-channel (left) and $t$-channel (right).}
    \label{fig:triangle_diangram}
\end{figure}

In this paper, we study the production of the pseudoscalar Higgs $A$ in the proposed muon collider within the Type-II and Type-X 2HDM. The results can be extrapolated to other models easily enough. In muon colliders, the production of $A$ at tree level is suppressed by the small Yukawa coupling, and as it does not couple to $W$ or $Z$ at tree level, the 1-loop contributions will be important. The production of $A$ that proceeds via the fusion of $\gamma\gamma$, $ZZ$, and $\gamma Z$ through a fermion loop is shown in Figure \ref{fig:triangle_diangram}, where the dominant contribution arises from $f = \{t, b, \tau \}$. Within the muon collider, the production of $A$ through the triangle diagrams proceeds via the $s$- (left) and $t$- (right) channels.
 

We compare the production cross section of $A$ when the contributions from the effective vertices are included through these triangle diagrams, with that at the tree level. For Type-II, we find that when not including the effective vertex contributions, the cross section ranges between $\sim 10^{-3} - 0.5$ fb at a COM energy of $\sqrt{s} = 3$ TeV for $\tan{\beta} \sim 5-40$ and $m_{A}$ up to 2 TeV. However, we find that the NLO contributions from the $\gamma\gamma$, $ZZ$, and $\gamma Z$ are comparable to the LO contribution and even larger in some regions of the parameter space, leading to an enhancement of up to $\sim 2$ in the low $m_A$ and low $\tan \beta$ region. We find similar results for Type-X, with the LO cross section ranging between $\sim 10^{-3} - 0.7$ fb for the same range of the parameters, and we also find the NLO to be also comparable or larger than the LO, with an enhancement factor that could reach $\sim 2-8$ in the low $m_A$ region, especially at low $\tan \beta$. We also study the impact of varying $\tan{\beta}$ on the cross section and find that the largest cross sections are generally obtained at high $\tan{\beta}$. However, we find that the strongest enhancement in both Type-II and Type-X happens at low $\tan \beta$, which is mainly due to the contribution from the top loop whose coupling $\sim \cot \beta$ in both types. We also investigate the impact of varying the COM energy on the cross section. We find that while the LO cross section exhibits the typical $1/s$ phase space suppression, the NLO becomes more important at high COM energies, which is a direct result of the derivative-type couplings of the effective vertices that yields energy enhancement that competes with the energy suppression from the phase space integration.

This paper is organized as follows: In Section \ref{sec:EFT}, we present the  EFT approach to the NLO corrections  and extract the effective vertices to be used in the simulation. In Section \ref{sec:results}, we present the simulation results, which include the impact of including the NLO corrections, the impact of $\tan{\beta}$ and the effect of $\sqrt{s}$ on the production cross section. We end with the conclusions and a discussion on future research possibilities in Section \ref{sec:outlook}.

\section{Electroweak NLO production cross sections: Effective Lagrangian}\label{sec:EFT}

In this section, we present our formalism for implementing the NLO corrections to the production of $A$. We will assume CP conservation throughout our paper. In our calculation, it is convenient to utilize an EFT approach by integrating out the fermion loops. Therefore, we write the effective Lagrangian as
\begin{equation}\label{eq:Eff_lag}
\mathcal{L}^{A}_{\text{eff}} \supset  \frac{g^{A}_{\gamma\gamma}}{v}AA_{\mu\nu}\widetilde{A}^{\mu\nu} + \frac{g^{A}_{\gamma Z}}{v}AA_{\mu\nu}\widetilde{Z}^{\mu\nu} + \frac{g^{A}_{ZZ}}{v}AZ_{\mu\nu}\widetilde{Z}^{\mu\nu},
\end{equation} 
where $V_{\mu\nu} = \partial_{\mu}V_{\nu} - \partial_{\nu}V_{\mu}$, $\widetilde{V}^{\mu\nu} = \frac{1}{2}\epsilon^{\mu\nu\rho\sigma}V_{\rho\sigma}$, and the VEV was introduced to keep the couplings dimensionless. This is reminiscent of the SM Higgs effective Lagrangian (see for instance \cite{Carmi:2012in}), however, we should keep in mind that unlike the SM Higgs, the CP-odd Higgs $A$ does not couple at tree level to gauge bosons, therefore there will be no gauge boson loops and only the fermion loops contribute. In our calculation, we only keep $f = \{t,b,\tau \}$ as they have the largest Yukawa couplings.
Starting with the $AVV$ vertex, we can write the effective interaction using the effective Lagrangian in eq. (\ref{eq:Eff_lag})
\begin{equation}\label{eq:A_gamma_gamma_vertex}
AV_{1}V_{2} \hspace{2mm}\text{vertex} = n\frac{2 i g_{V_{1}V_{2}}^{A}}{v}k_{1\rho}k_{2\sigma}\epsilon^{\mu\nu\rho\sigma},
\end{equation}
where $k_{1,2}$ are the momenta of the gauge bosons, and $n=1 (2)$ for $V_{1} \neq V_{2}$ $(V_{1} = V_{2})$. Using this EFT vertex, it is straightforward to calculate the decay width $A \rightarrow \gamma\gamma$,
\begin{equation}\label{eq:A_decayWidth_EFT}
\Gamma_{\text{EFT}}(A \rightarrow \gamma\gamma) = \frac{(g_{\gamma \gamma}^{{A}})^{2}}{4\pi v^{2}}M_{A}^{3}.
\end{equation}
On the other hand, the exact results are given by \cite{Gunion:1989we}
\begin{equation}\label{eq:AdecayWidth_Exact}
\Gamma_{\text{Exact}}(A \rightarrow \gamma\gamma) = \frac{\alpha^{2}g^{2}}{1024\pi^{3}}\frac{M_{A}^{3}}{M_{W}^{2}}\Big|\sum_{f}N_{c}^f Q_{f}^{2}R_{f} F_{1/2}^{A}(\tau_{f}) \Big|^{2},
\end{equation}
where $N_{c}^{f}$ is the color factor, $R_{f}$ is the ratio of the Yukawa couplings of $A$ relative to the SM Higgs given in Table \ref{tab:Yukawa_2HDM}, $\tau_{f} = \frac{4m_{f}^{2}}{M_{A}^{2}}$, and the function $F_{1/2}^{A}(\tau)$ is given by
\begin{equation}\label{eq:F1/2}
F_{1/2}^{A}(\tau)  = -2\tau f(\tau),
\end{equation}
with
\begin{equation}\label{eq:f(tau)}
f(\tau) = \begin{cases} 
     \Big[ \sin^{-1}(1/\sqrt{\tau})\Big]^{2}  & \tau \geq 1, \\
      -\frac{1}{4} \Big[ \log{\Big( \frac{1+\sqrt{1-\tau}}{1-\sqrt{1-\tau}}\Big) - i\pi)}\Big]^{2} & \tau <  1. \\
   \end{cases}
\end{equation}
Thus, we can easily find $g_{\gamma \gamma}^{A}$ by simply matching eq. (\ref{eq:A_decayWidth_EFT}) to eq. (\ref{eq:AdecayWidth_Exact}). We obtain
\begin{equation}\label{eq:EFT_A_gamma}
g_{\gamma\gamma}^{A} = \frac{\alpha}{8\pi} \sum_{f}N_{c}^{f}Q_{f}^{2}R_{f} F^{A}_{1/2}(\tau_{f}).
\end{equation}

Similarly, it is easy to obtain $g_{\gamma Z}^{A}$ by calculating the decay width from the effective Lagrangian (\ref{eq:Eff_lag}) then matching it to the exact result, which can be found for instance in \cite{Gunion:1989we}. We obtain, 
\begin{equation}\label{eq:EFT_A_gammaZ}
    g_{\gamma Z}^{A} = \frac{\alpha}{4\pi}\sum_{f}2N^{f}_{c}Q_{f}R_{f}\frac{(I_{f}^{3}-2Q_{f}\sin^{2}{\theta_{W}})}{\sin{\theta_{W}}\cos{\theta_{W}}}I_{2}(\tau_{f},\lambda_{f}),
\end{equation}
where $\theta_{W}$ is the Weinberg angle, $\tau_{f} = \frac{4m_{f}^{2}}{M_{A}^{2}}$, $\lambda_{f} = \frac{4m_{f}^{2}}{M_{Z}^{2}}$, $I_{f}^3$ is the third component of weak isospin for the fermion $f$ and the the function $I_{2}$ is given by
\begin{equation}
    I_{2}(\tau,\lambda) = \frac{-\tau \lambda}{2(\tau - \lambda)}[f(\tau) - f(\lambda)].
\end{equation}

Finally, we turn our attention to $g^{A}_{ZZ}$. This effective coupling was calculated in \cite{Gunion:1991cw, Aiko:2022gmz}, and we find it more convenient to use their results and formalism instead of that in eq. (\ref{eq:Eff_lag}), although matching their results to the effective Lagrangian is straightforward. In the notation of \cite{Aiko:2022gmz}, the effective $AZZ$ vertex in the case CP is conserved can be expressed as
\begin{equation}\label{eq:AZZ_eff}
    \widehat{\Gamma}^{\mu\nu}_{AZZ}(p_{1},p_{2},q) = i\epsilon^{\mu\nu\rho\sigma} p_{1\rho}p_{2\sigma} \times \frac{i g_{Z}^{2}}{16\pi^{2}}\sum_{f}\frac{16I_{f}N_{c}^{f}m_{f}^{2}R_{f}}{v}\Big[v_{f}^{2}C_{0} + a_{f}^{2}(C_{0} + 2C_{11} - C_{12})\Big](f,f,f),
\end{equation}
where $g_{Z}^{2} = g^{2} + g^{\prime 2}$, $v_{f} = \frac{1}{2}I_{f} - Q_{f} \sin^{2}{\theta_{W}}$, $a_{f} = \frac{1}{2}I_{f}$ and $C_{0}, C_{11}$ and $C_{12}$ are the three-point Passarino-Veltman functions \cite{Passarino:1978jh, Ellis:2011cr}, which can be expressed as follows: 
\begin{align}
C_0(M_Z^2, M_Z^2, M_A^2 , m_f^2, m_f^2, m_f^2)  = &
\frac{1}{M_A^2 - 4m_f^2} 
  \Big[B_0(M_Z^2,m_f^2,m_f^2) - B_0(M_A^2,m_f^2,m_f^2) \Big],
\\
  C_{11}(M_Z^2, M_Z^2, M_A^2, m_f^2, m_f^2, m_f^2)  = & \; \frac{ M_A^2+ 3M_Z^2}{M_A^2 - 4 M_Z^2} \;  C_0(M_Z^2, M_Z^2, M_A^2, m_f^2, m_f^2, m_f^2), \\
  C_{12}(M_Z^2, M_Z^2, M_A^2, m_f^2, m_f^2, m_f^2)  = & \; \frac{M_Z^2 }{M_A^2 - 4 M_Z^2} \; C_0(M_Z^2, M_Z^2, M_A^2, m_f^2, m_f^2, m_f^2),
\end{align}
where $B_0$ is the two-point Passarino-Veltman function, which can be expressed in terms of the masses as follows:
\begin{equation}
B_0(M_A^2, M_f^2, m_f^2) = \begin{cases}
    \Delta_{\epsilon} + 2 - 2 \log\left( \frac{m_f^2}{\mu^2} \right)
+ \frac{2 \sqrt{1 - \frac{4m_f^2}{M_A^2}}}{M_A^2} 
\log\left( \frac{1 + \sqrt{1 - \frac{4m_f^2}{M_A^2}}}{1 - \sqrt{1 - \frac{4m_f^2}{M_A^2}}} \right) &  m_{f} > M_{Z}, \\
\Delta_{\epsilon} + 2 - 2 \log\left( \frac{m_f^2}{\mu^2} \right)
+ \frac{2\sqrt{\frac{4m_f^2}{M_Z^2} - 1}}{M_Z^2} \tan^{-1}\left( \frac{1}{\sqrt{\frac{4m_f^2}{M_Z^2} - 1}} \right), & m_{f} < M_{Z},
\end{cases}    
\end{equation}
where $\mu$ is the renormalization scale, and $\Delta_{\epsilon}$ represents the dimensional regularization divergent part, which cancels in $C_0$, leaving it as a finite quantity.

Having the mathematical expressions in hand, we wish to emphasize a few points about the effective vertices. All three of these vertices are directly proportional to $R_f$, the ratio of the Yukawa couplings as given in Table \ref{tab:Yukawa_2HDM}. If one concentrates only on the coupling to quarks, then there are two classes of 2HDMs. In class A models, consisting of Type-I and Type-X, both $R_u^A$ and $R_d^A$ are given by $\cot\beta$, which implies significant enhancement of the cross section at low $\tan\beta$. On the other hand, for class B models consisting of the remaining two models (Type-II and Type-Y), we can see $R_d^A \sim \tan\beta$, whereas $R_u^A \sim \cot\beta$. Thus, only up-type quarks (in our case, the top quark), will lead to enhancement at low $\tan\beta$. This implies that for class B models, there will be a competition between the enhancement (suppression) due to the top quark and the suppression (enhancement) due to the bottom quark at low (high) $\tan\beta$. A more detailed analysis for distinguishing class A models with those of class B is thus possible and we intend to do it in the future.
\section{Results}\label{sec:results}
In this section, we present our analysis for the production of $A$ in muon colliders within the Type-II and Type-X 2HDM. We implemented the effective vertices introduced above, namely $\gamma\gamma A$, $\gamma Z A$ and $ZZA$; in \texttt{FeynRules} \cite{Alloul:2013bka} to generate the \texttt{UFO} files \cite{Degrande:2011ua} which were used in \texttt{MadGraph5\char`_aMC@NLO} \cite{Alwall:2011uj} to generate $\mu^{+}\mu^{-} \rightarrow \mu^{+}\mu^{-} A$ events. To limit the background in our simulation, we require a transverse momentum cut of $p_T^{\ell} \geq 10$ GeV for the muons, and require their pseudorapidity to be $|\eta^{\ell}| \leq 3.5$. We also require the final state muons be within $\Delta R \geq 0.4$.

\begin{figure}[t!]
    \centering
    \includegraphics[width=7.9cm,height=6.5cm]{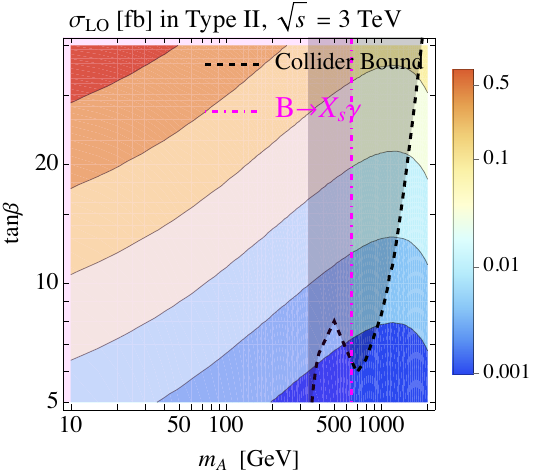}
\includegraphics[width=7.9cm,height=6.5cm]{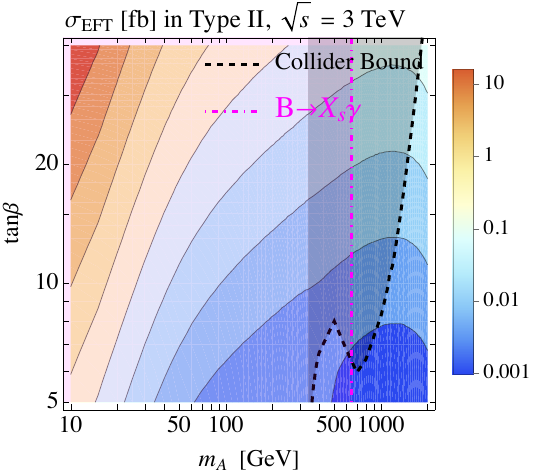}
\caption{The cross section for the process $\mu^+\mu^- \rightarrow \mu^+\mu^-A$ in the $m_A \, - \, \tan\beta$ parameter space for Type-II. In the left plot, we show the cross section at tree level, whereas on the right plot, we include the NLO corrections through the effective vertices. We fix $\sqrt{s} = 3$ TeV. The black-dashed contour displays the excluded region by the LHC \cite{ATLAS:2020zms}, whereas the magenta dashed-dot line shows the excluded region from the $Br(B\rightarrow X_s\gamma)$ decay \cite{Misiak:2017bgg}.}
    \label{fig:Combined_plot}
\end{figure}

\begin{figure}[t!]
    \centering
    \includegraphics[width=7.9cm,height=6.8cm]{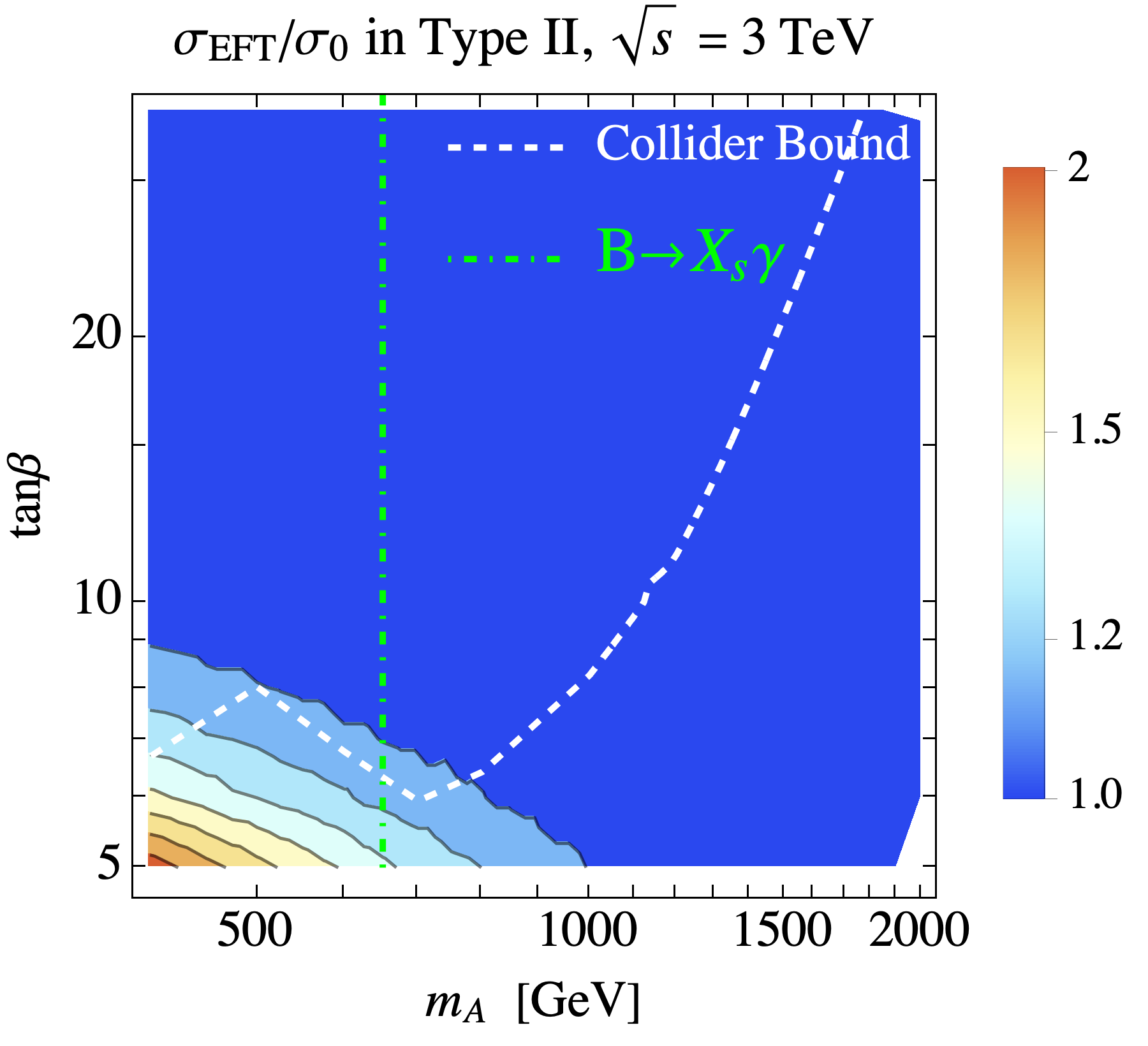}
\caption{The ratio of $\sigma_{\text{EFT}}$ to $\sigma_{0}$ for Type-II after including $\gamma\gamma$, $\gamma Z$, and $ZZ$ effective vertices at $\sqrt{s} = 3$ TeV. The white-dashed contour displays the excluded region by the LHC \cite{ATLAS:2020zms}, whereas the green dashed-dot line shows the excluded region from the decay $B\rightarrow X_s\gamma$\cite{Misiak:2017bgg, Belle:2016ufb, ParticleDataGroup:2024cfk}.}
    \label{fig:ratio_plot}
\end{figure}

We first investigate the impact of the NLO corrections on the cross section of $\mu\mu \rightarrow \mu\mu A$. We fix the COM energy to $\sqrt{s} = 3$ TeV and vary $\tan{\beta}$ between 5 and 40 in order to avoid the breakdown of perturbativity in the couplings of $b$ and $\tau$ ($\tau$) at very high $\tan{\beta}$ for Type-II (Type-X). The results are shown in Figure \ref{fig:Combined_plot} in the $m_{A}-\tan{\beta}$ parameter space. The left plot shows the tree level cross section alone, whereas the right plot shows the cross section after including the NLO corrections through the effective vertices. We can see from the plots that while the tree level cross section is mostly suppressed, ranging between $\sim 0.001 - 0.5$ fb, including the NLO could enhance the cross section up to $\sim 10$ fb in certain regions of the parameter space. However, this enhancement is not uniform, and as we shall see, the enhancement represents a factor of $\sim 1-2$ depending on the region of the parameter space. This clearly illustrates how the contributions of the loop diagrams are at least as important as the ones at tree level, if not more significant, thereby demonstrating the necessity of including the NLO corrections. 

Another interesting point to consider is the impact of $\tan{\beta}$ on the cross section. When confining the cross section to the tree level, it is quite obvious that there is a direct relationship between $\tan{\beta}$ and the cross section, such that a larger $\tan{\beta}$ always increases the cross section. This can be understood by inspecting Table \ref{tab:Yukawa_2HDM}, where we see that $R_{\mu}^{A} = \tan{\beta}$, which explains the tree-level enhancement. Things are slightly different when including the NLO corrections, where we see that there is noticeable enhancement at low $\tan{\beta}$ (which is more clearly visible in Figure \ref{fig:ratio_plot}). Once again this can be understood by inspecting Table \ref{tab:Yukawa_2HDM}. As we are only including $f =\{t,b,\tau\}$, and for Type-II $R^{A}_{t} = \cot \beta$ and $R^{A}_{b,\tau} = \tan \beta$, we can see that the enhancement at low $\tan{\beta}$ is mainly due to the top loop, which demonstrates its dominance over the other two loops.

We superimpose the latest bounds from the decay $B \rightarrow X_{s}\gamma$ \cite{Misiak:2017bgg, Belle:2016ufb, ParticleDataGroup:2024cfk} and from the LHC searches \cite{ATLAS:2020zms, ParticleDataGroup:2024cfk} on the plots. We can see that the former searches exclude all of the region $m_{A} \lesssim 700$ GeV, whereas the latter excludes much of the parameter space for $m_{A} \lesssim 1500$ GeV, especially for high $\tan\beta$, although some region is still open. Ignoring the region excluded by experiment, we can see that in total, cross sections up to $\sim 1$ fb are achievable, especially at $m_{A} \gtrsim 1500$ GeV and high $\tan{\beta}$. 

To further illustrate the significance of the NLO contribution, we plot the ratio of the cross section including the EFT corrections to that at tree level in the $\tan{\beta}-m_{A}$ parameter space in Figure \ref{fig:ratio_plot}. There, we can see that the NLO contribution is as large as the tree level contribution, and can be even larger for low $m_{A}$ and low $\tan{\beta}$, with a potential enhancement reaching a factor of $\sim 2$ (although the highest enhancement coincides with the region excluded by experiment). We can also clearly see the aforementioned dominance of the top loop by inspecting the enhancement in the low $\tan{\beta}$ region.

\begin{figure}[t!]
    \centering
    \includegraphics[width=7.9cm,height=6.5cm]{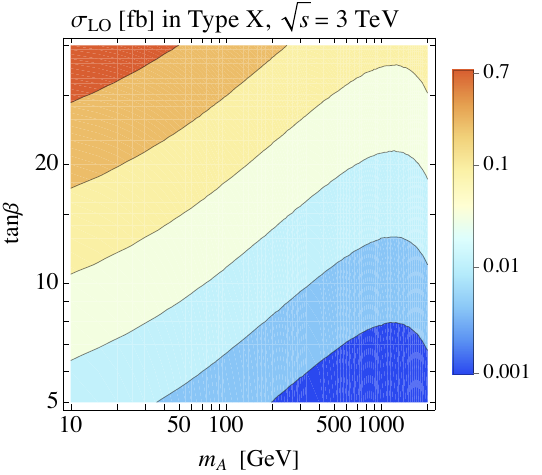}
\includegraphics[width=7.9cm,height=6.5cm]{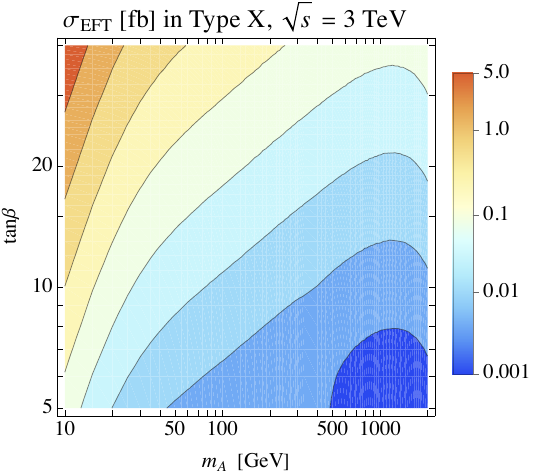}
\caption{The cross section for the process $\mu^+\mu^- \rightarrow \mu^+\mu^-A$ in the $m_A \, - \, \tan\beta$ parameter space for Type-X. In the left plot, we show the cross section at tree level, whereas on the right plot, we include the NLO corrections through the effective vertices. We fix $\sqrt{s} = 3$ TeV. For Type-X, there is no excluded region.}
    \label{fig:Combined_plot_Type_X}
\end{figure}

\begin{figure}[b!]
    \centering
    \includegraphics[width=7.9cm,height=6.8cm]{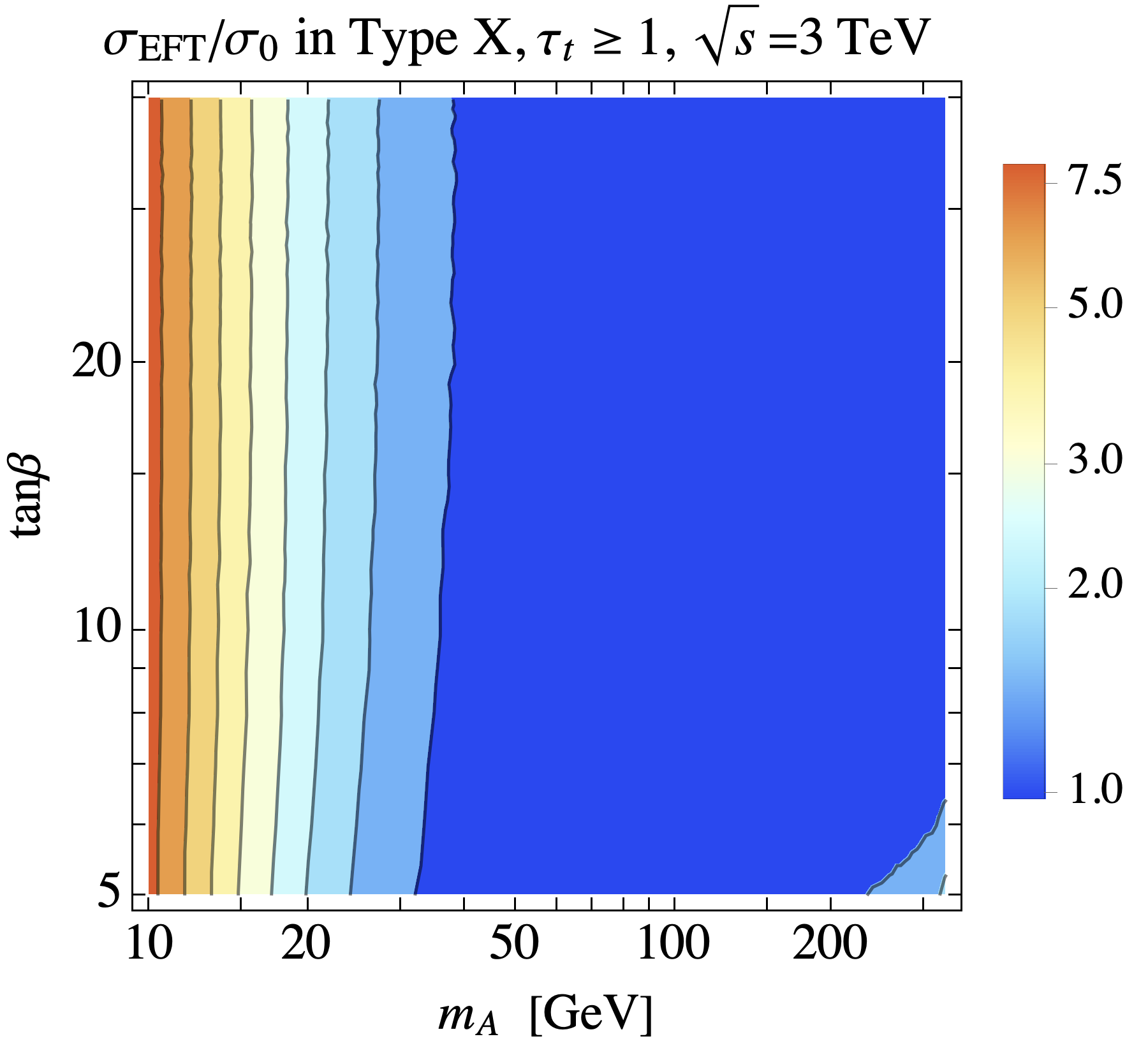}
    \includegraphics[width=7.9cm,height=6.8cm]{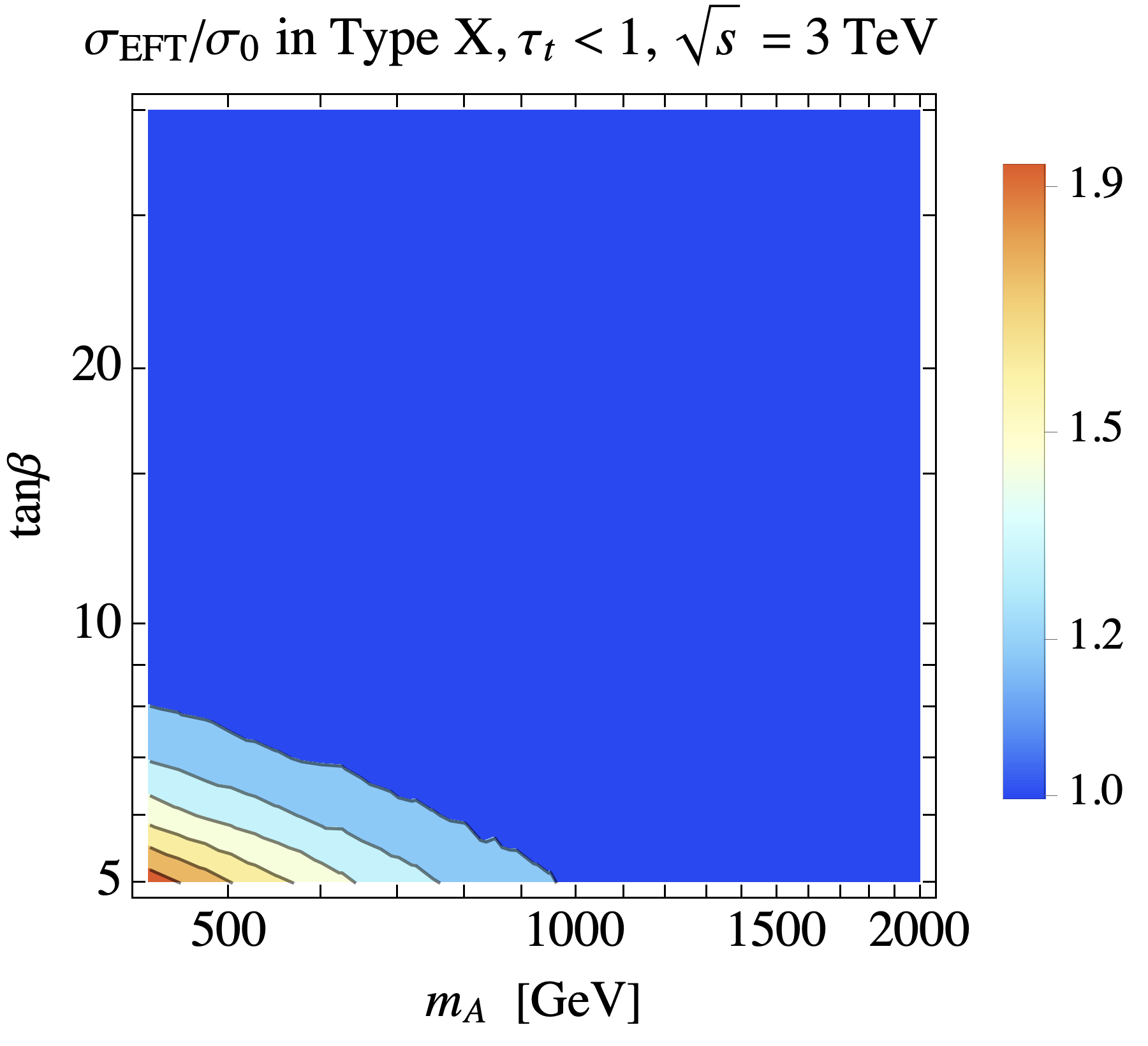}
\caption{The ratio of $\sigma_{\text{EFT}}$ to $\sigma_{0}$ for Type-X after including $\gamma\gamma$, $\gamma Z$, and $ZZ$ effective vertices at $\sqrt{s} = 3$ TeV for $\tau_{t} \geq 1$ (left) and $\tau_{t} < 1$ (right). For Type-X, there is no excluded region.}
    \label{fig:ratio_plot_type_X}
\end{figure}

\begin{figure}[t!]
    \centering
    \includegraphics[width=7.9cm,height=6.5cm]{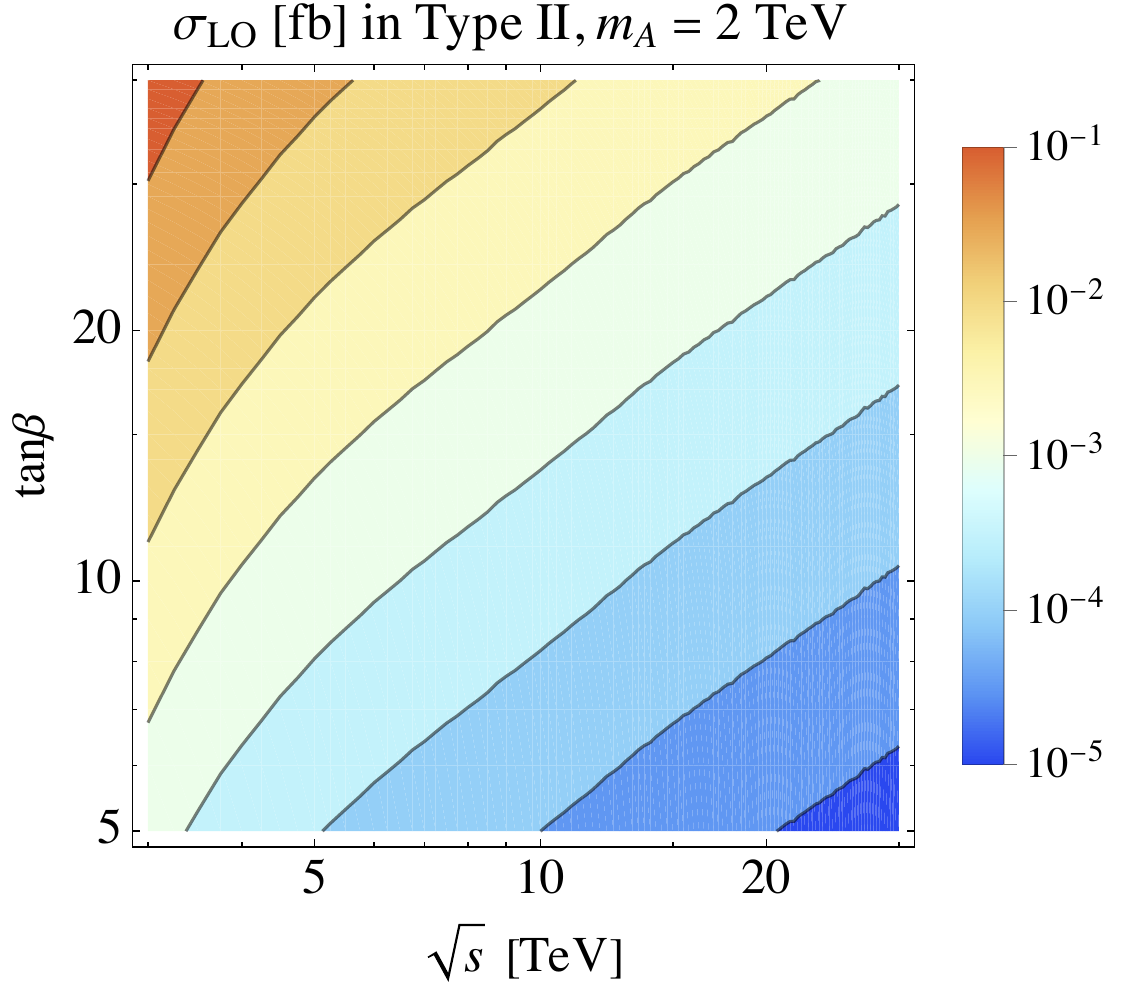}
     \includegraphics[width=7.9cm,height=6.5cm]{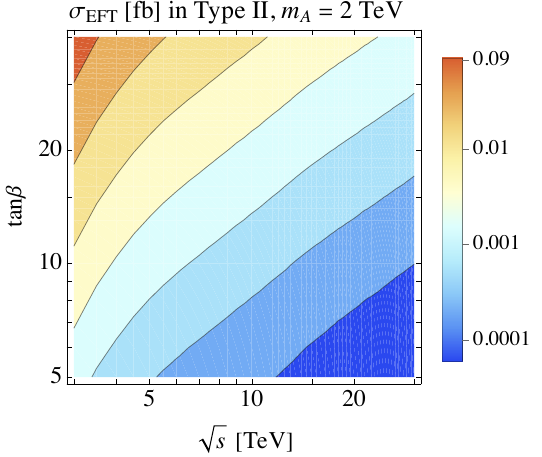}
\caption{The cross section for the process $\mu^+\mu^- \rightarrow \mu^+\mu^-A$ in the $\sqrt{s} \, - \, \tan\beta$ parameter space for Type-II. In the left plot, we show the cross section at tree level, whereas on the right plot, we include the NLO corrections through the effective vertices. We fix $m_{A} = 2$ TeV.}
    \label{fig:sqrt_s_vary}
\end{figure}

\begin{figure}[b!]
    \centering
     \includegraphics[width=7.9cm,height=6.5cm]{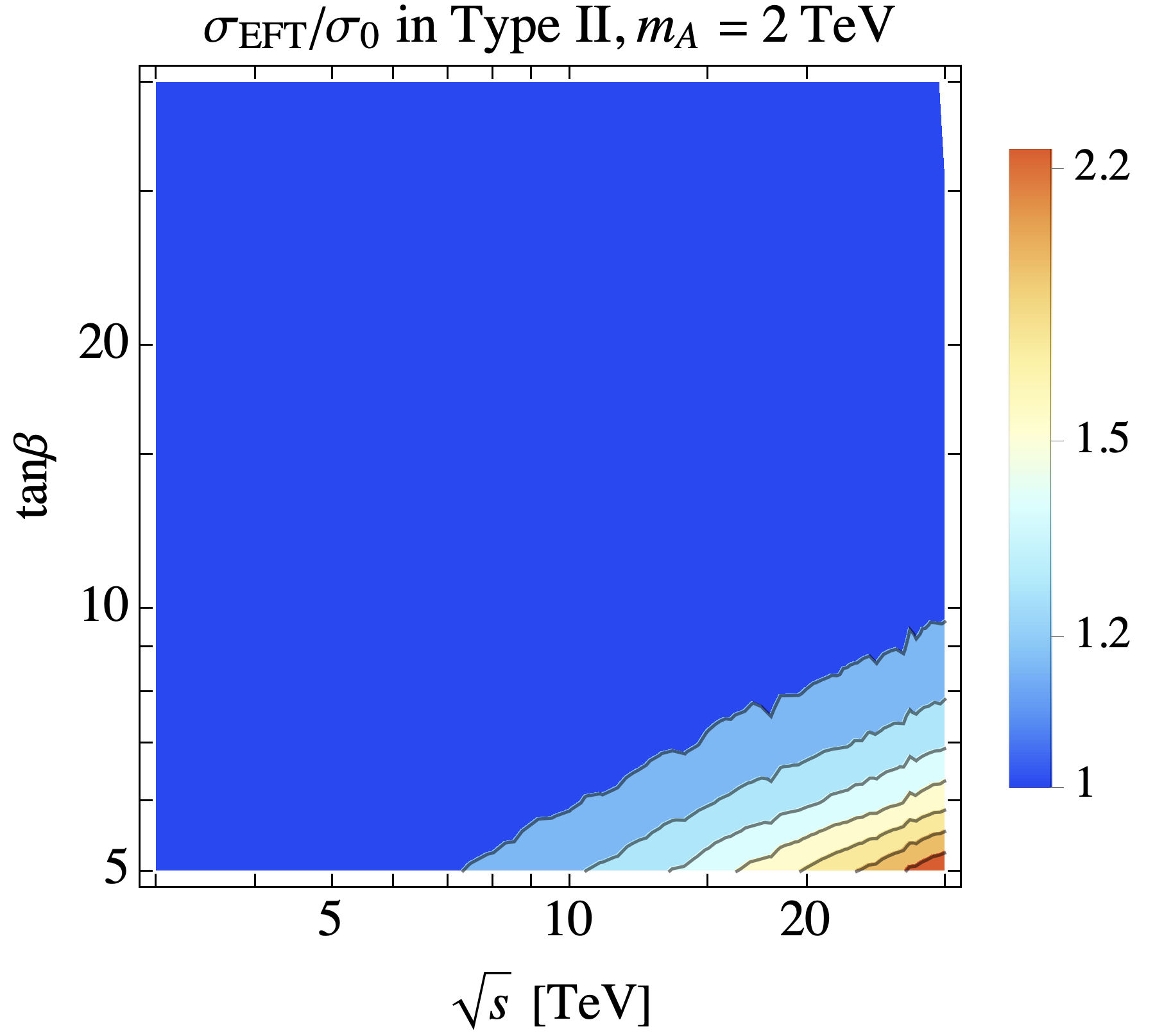}
\caption{The ratio of $\sigma_{\text{EFT}}$ to $\sigma_{0}$ for Type-X after including the $\gamma\gamma$, $\gamma Z$, and $ZZ$ effective vertices in the $\tan\beta - \sqrt{s}$ parameter space. Here we fix $m_A = 2$ TeV.}
    \label{fig:sqrt_s_vary_ratio}
\end{figure}

\begin{figure}[t!]
    \centering
    \includegraphics[width=7.9cm,height=6.5cm]{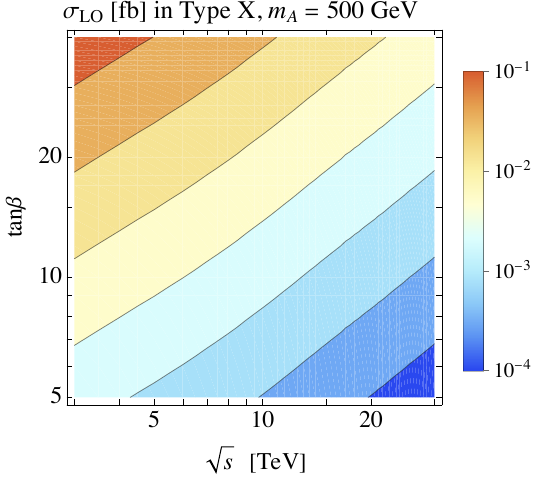}
     \includegraphics[width=7.9cm,height=6.5cm]{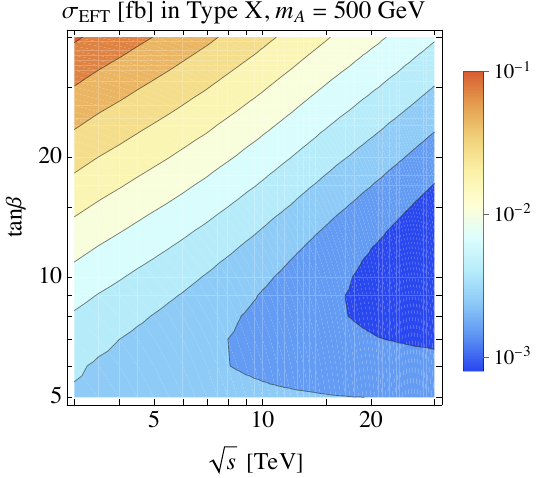}
     \includegraphics[width=7.9cm,height=6.5cm]{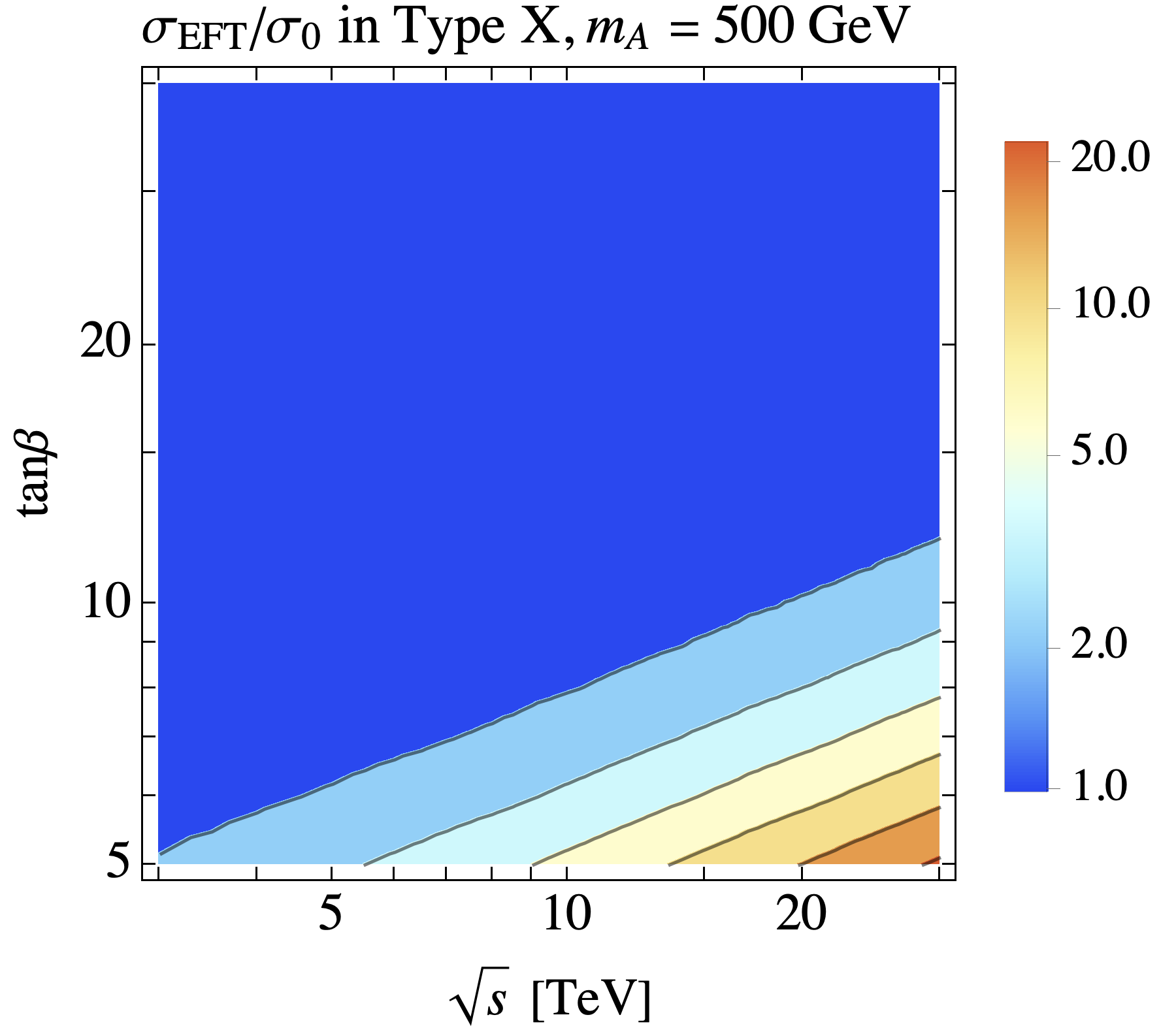}
\caption{The cross section for the process $\mu^+\mu^- \rightarrow \mu^+\mu^-A$ in $\sqrt{s} \, - \, \tan\beta$ parameter space for Type-X at tree level (top left) and after including the $\gamma\gamma$, $\gamma Z$, and $ZZ$ effective vertices (top right) in the $\tan\beta - \sqrt{s}$ parameter space. (Bottom) The ratio of $\sigma_{\text{EFT}}$ to $\sigma_{0}$. Here we fix $m_A = 500$ GeV.}
    \label{fig:sqrt_s_vary_type_X}
\end{figure}

Turning our attention to the Type-X model, we redo the same analysis and show the results in Figure \ref{fig:Combined_plot_Type_X}, where we see a similar behavior. More specifically, the LO cross section ranges between $\sim 0.001-0.7$ fb, but could potentially increase to $\sim 5$ fb when including the NLO corrections. Similar to the Type-II 2HDM, the cross section, whether at tree level or at 1-loop, is largest in the low $m_A$ high $\tan\beta$ region, and suppressed in the high $m_A$ low $\tan\beta$ region, but with the relative enhancement of the NLO to the LO contribution arising mostly in the low $\tan \beta$ regime. This relative enhancement at low $\tan\beta$ demonstrates the dominance of the top loop. In addition, the low $\tan\beta$ region receives additional enhancement from the bottom loop, since in Type-X $R^{A}_{b} = \cot \beta$. An important difference between the Type-II and Type-X models is that, unlike the former, the experimental bounds on the latter are weak, and thus the parameter space is essentially unconstrained \cite{Das:2024ekt, Jueid:2021avn, Ma:2024ayr}.

We also plot the ratio of NLO to LO cross sections in Figure \ref{fig:ratio_plot_type_X}. There, we plot the regions where $\tau_{t} \geq 1$ and $\tau_{t} < 1$ separately.\footnote{We did not plot the region $\tau_{t} \geq 1$ for the Type-II model since that region is excluded by experiment.} For $\tau_{t} \geq 1$ shown on the LHS, we see that the NLO contribution could potentially lead to an order of magnitude enhancement, especially at low $m_A$. We also see that $\tan \beta$ has little impact on the ratio in that regime. On the other hand, for $\tau_{t} < 1$, the enchantment could reach a factor of $\sim 2$, especially at the low $m_A$ low $\tan \beta$ region. The low $\tan \beta$ enhancement for $\tau_{t} < 1$ can easily be understood as being a direct result of the dominance of the top loop. Things are a bit more complicated for $\tau_{t} > 1$ and can be understood by inspecting eq. (\ref{eq:f(tau)}): In the region $300 \gtrsim m_A \gtrsim10$ GeV, we have $\tau_{t}> 1$ and $\tau_{b,\tau} < 1$. In that regime $f_{\tau_{t}}$ is negligible, whereas $f_{\tau_{b}}$ and $f_{\tau_{\tau}}$ are comparable to one another, and since for Type-X, $R^{A}_{b} = -\cot \beta$ and $R^{A}_{\tau} = \tan \beta$, there is a competition between the two contributions that essentially leads to their offsetting one another, which explains the minimal impact of $\tan \beta$ in that regime.

So far we have fixed $\sqrt{s}$ to 3 TeV, however, as the proposed COM energy for the future collider varies, with suggested energies ranging between 3 and 30 TeV, it is worthwhile investigating the impact of $\sqrt{s}$ on the production cross section of $A$. Starting with Type-II, we fix $m_{A} = 2$ TeV to avoid the experimental bounds, and use the same cuts above. In Figure \ref{fig:sqrt_s_vary}, we show the cross section of $\mu\mu \rightarrow \mu\mu A$ in the $\tan{\beta}-\sqrt{s}$ parameter space, at tree level (left) and after including the EFT vertices (right). First, we notice the usual behavior where the NLO contribution is comparable to the tree level one, as observed above. Second, we can see that both at tree level and at NLO, the production cross section drops with $\sqrt{s}$ at fixed $\tan{\beta}$, which reveals the $1/s$ dependence of the cross section arising from the phase space integration. 

\begin{figure}[t!]
    \centering
    \includegraphics[width=7.9cm,height=6.5cm]{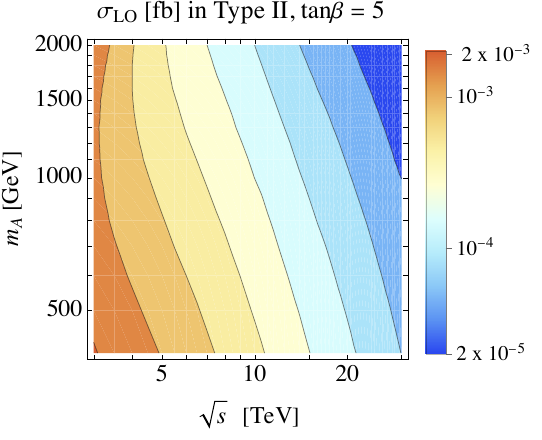}
    \includegraphics[width=7.9cm,height=6.5cm]{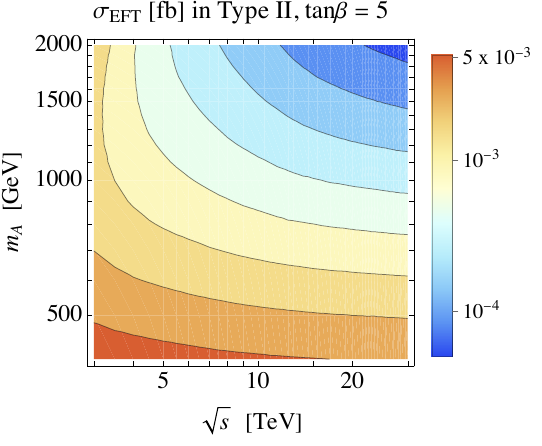}
    \includegraphics[width=7.9cm,height=6.5cm]{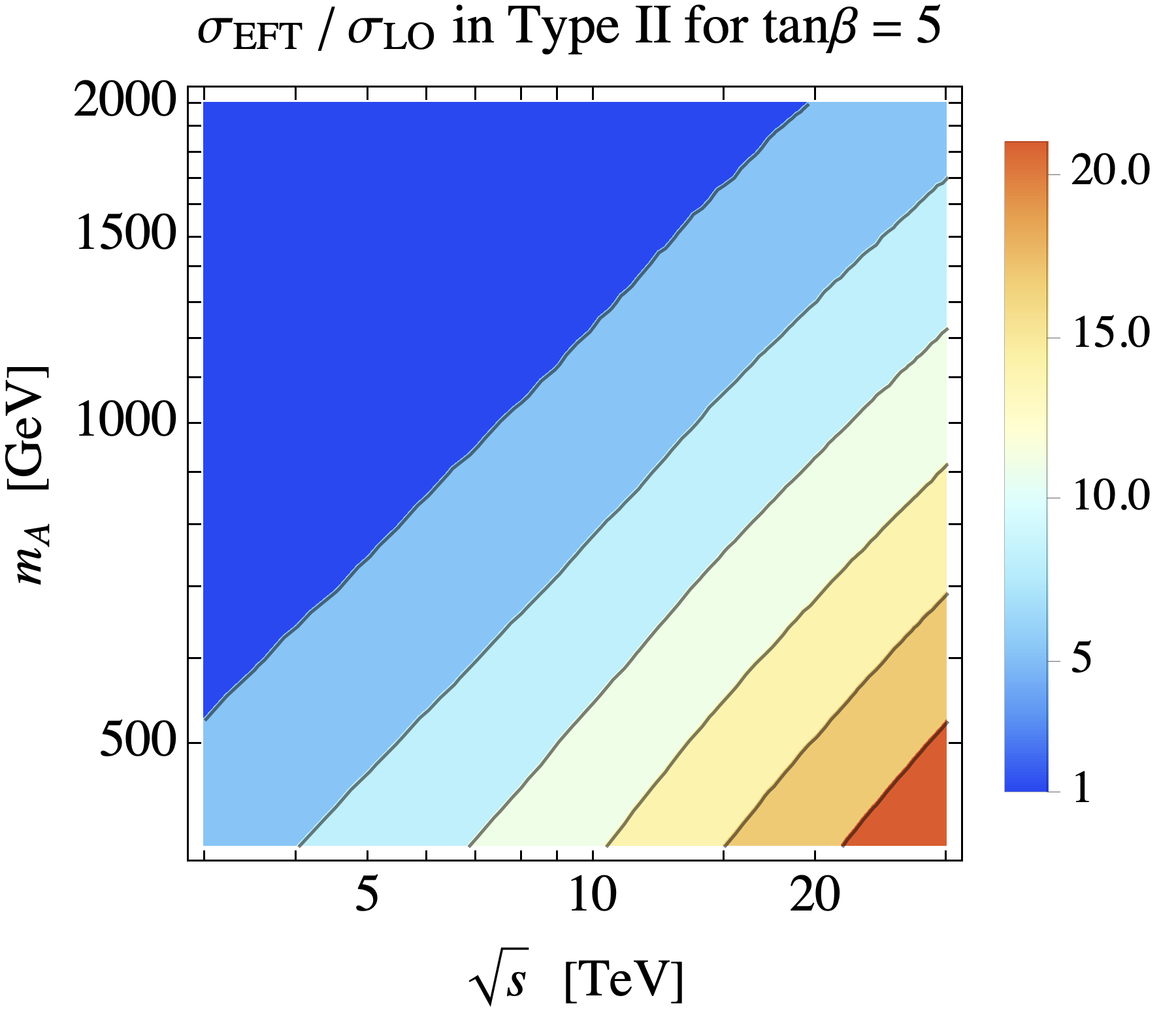}
\caption{The production cross section for the process $\mu^+\mu^- \rightarrow \mu^+\mu^-A$ in $\sqrt{s} \, - \, m_A$ parameter space at LO (top left) and at NLO (top right) for Type-II. (Bottom) The ratio $\sigma_{\text{EFT}}/\sigma_{0}$. Here we fix $\tan{\beta} = 5$.}
    \label{fig:energy_mA_Type_II}
\end{figure}

\begin{figure}[t!]
    \centering
    \includegraphics[width=7.9cm,height=6.5cm]{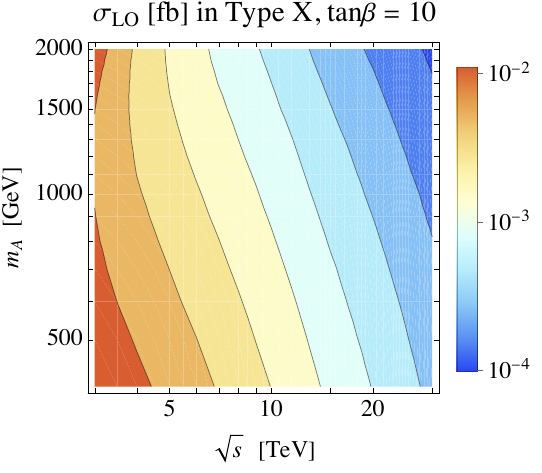}
     \includegraphics[width=7.9cm,height=6.5cm]{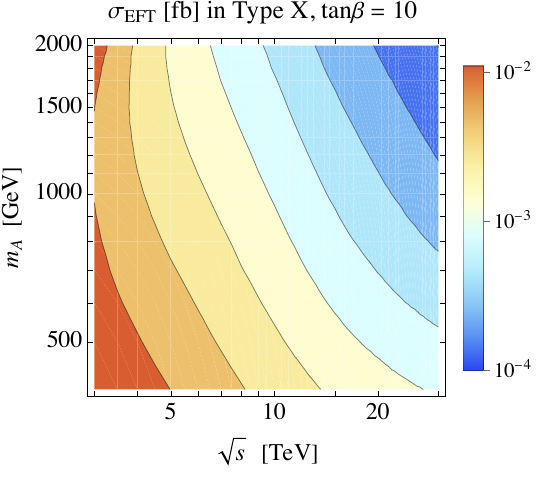}
    \includegraphics[width=7.9cm,height=6.5cm]{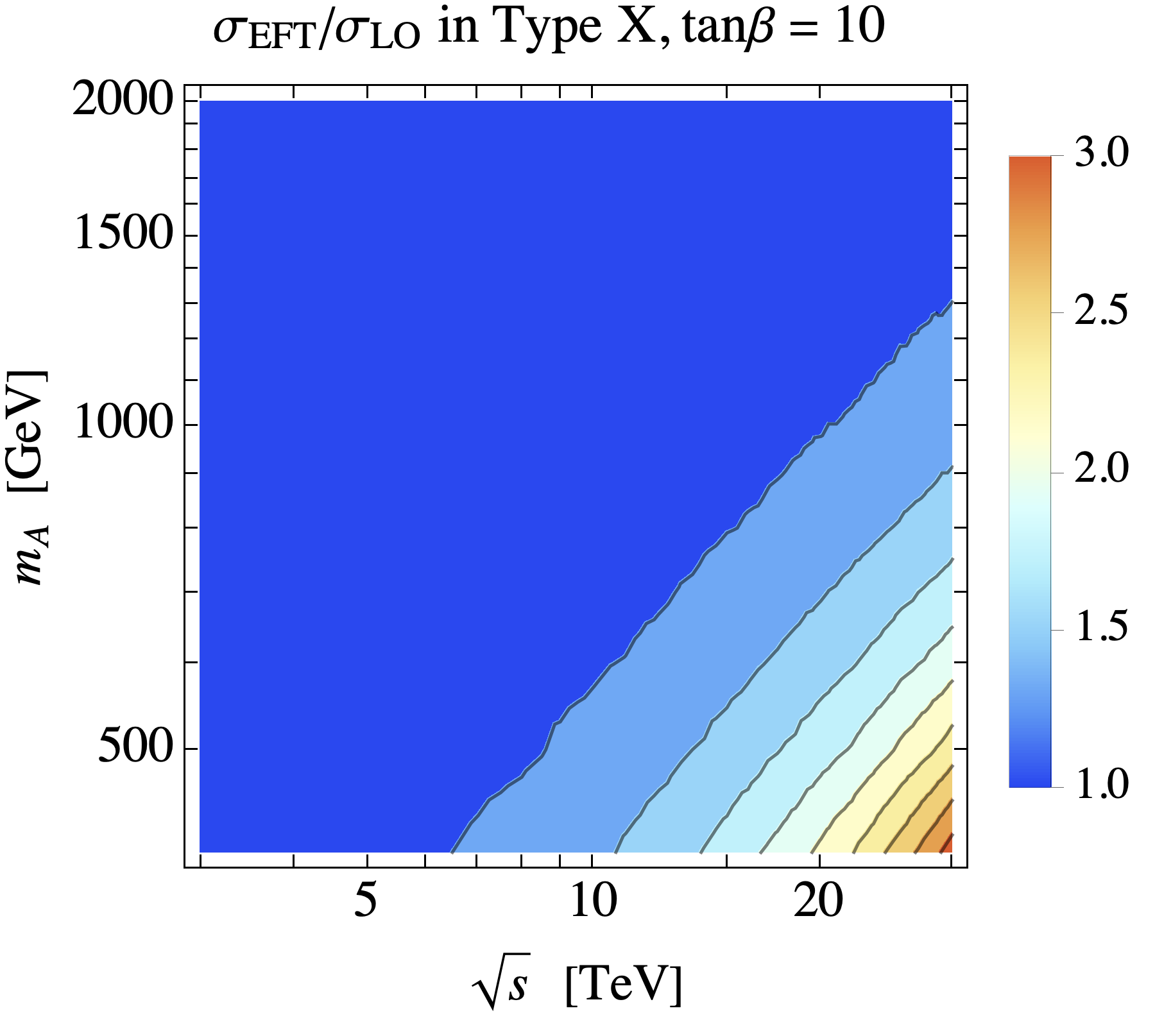}
\caption{The production cross section for the process $\mu^+\mu^- \rightarrow \mu^+\mu^-A$ in $\sqrt{s} \, - \, m_A$ parameter space at LO (top left) and at NLO (top right) for Type-X. (Bottom) The ratio $\sigma_{\text{EFT}}/\sigma_{0}$. Here we fix $\tan{\beta} = 10$}
    \label{fig:energy_mA_Type_X}
\end{figure}

We show the impact $\sqrt{s}$ has on enhancing the NLO contribution over the LO contribution more explicitly in Figure \ref{fig:sqrt_s_vary_ratio}, where we plot the ratio of $\sigma_{\text{EFT}}/\sigma_{0}$. As the figure shows, for most of the parameter space, $\sqrt{s}$ has little impact on the enhancement of the cross section, except for low $\tan \beta$ and at high energies $\sqrt{s} \gtrsim 7$ TeV, where the enhancement could reach a factor of $\sim 2$. This can also be seen by comparing the bottom right corners of the plots in Figure \ref{fig:sqrt_s_vary}, where we see the NLO is enhanced in that region. The reason behind this arises from the derivative-type coupling (see eq. (\ref{eq:A_gamma_gamma_vertex})), which leads to energy enhancement that compensates for the energy suppression due to the phase space integration. This is most apparent at low $\tan \beta$ where the top loop dominates.

We repeat the same analysis for Type-X, but here, we set $m_A = 500$ GeV since the experimental bounds on Type-X are much weaker than on Type-II, and this mass window is still open. In the top row of Figure (\ref{fig:sqrt_s_vary_type_X}) we show the cross sections at LO and NLO. At LO, we see a similar behavior compared to Type-II, however, the behavior is slightly different at NLO: Although the behavior is quite similar to Type-II in most of the parameter space, it is somewhat different at high energy, where now the cross section is suppressed at intermediate $\tan \beta$, as opposed to low $\tan \beta$ in Type-II. More specifically, there is enhancement in the low $\tan \beta$ region in Type-X that does not exist in Type-II. This can be understood from Table \ref{tab:Yukawa_2HDM}, where we see than in Type-X, there is an additional enhancement at low $\tan \beta$ from the bottom loop since $R_{b}^{A} = \cot \beta$, which leads to this behavior. On the other hand, in Type-II $R_{b}^{A} = \tan \beta$ which leads to a suppression. 

In the bottom row of Figure \ref{fig:sqrt_s_vary_type_X} we show the ratio of the NLO cross section to the LO cross section, where we see similar behavior to Type-II, namely that there is enhancement at high energy and low $\tan \beta$, however, here the enhancement is much higher, reaching a factor of $\sim 20$, and kicks in at lower $\sqrt{s}$. This is due to the additional enhancement of the bottom loop in Type-X as discussed above. On the other hand, for both models $R^{A}_{\tau} = \tan\beta$ which is yields a suppressed contribution in that region. 

Finally, we investigate the impact of the COM energy on the production of ${A}$ when we vary $m_{A}$. We start with Type-II and fix $\tan{\beta} = 5$. Then we show the cross sections at LO (top left) and NLO (top right) and the ratio of the latter to the former (bottom) in Figure \ref{fig:energy_mA_Type_II}. The top plots reveal the expected $1/s$ suppression of the cross section as discussed, and the bottom plot shows the enhancement due to the NLO contribution, which is most significant in the high energy low $m_A$ region, and could reach a factor of $\sim 20$. We perform the same analysis for Type-X but with $\tan \beta =10$, and show the results in Figure \ref{fig:energy_mA_Type_X} where we observe a very similar behavior, except that the enhancement is lower, reaching $\sim 3$ at most.

We note in passing, that at high-energy muon colliders, the associated production of $A$ with of $t\bar{t}$ or $b\bar{b}$ is important. However, in the context of EW NLO corrections, the only new contribution will arise from the s-channel, as shown in Figure \ref{fig:triangle_diangram}, and as this channel is subdominant compared to the t-channel, the enhancement compared to the tree level will be negligible. We have verified that the NLO enhancement is insignificant.

We also point out that in muon colliders (as the case in $ee$ colliders), QCD corrections are not as important as in $pp$ colliders where the dominant production process is gluon fusion. In our case, QCD corrections arise at NNLO and are thus subleading. On the other hand, in $pp$ colliders, QCD corrections can be sizable and cannot be neglected. See for instance \cite{Bagnaschi:2022dqz}.

\section{Conclusion and Outlook}\label{sec:outlook}
In this paper, we investigated the impact of the NLO corrections to the production of the pseudoscalar $A$ within the Type-II and Type-X 2HDM in future muon colliders. In our analysis, we included the $\gamma\gamma A$, $\gamma Z A$, and $ZZ A$ triangle diagrams, keeping only ${\tau, b, t}$ in the loop. We found that the NLO contributions can be comparable to the tree level cross section, and even larger in certain regions of the parameter space. The reason is mainly due to the smallness of the Yukawa coupling of the muon, which leads to a suppression at tree level, thereby making the NLO contributions important.

Additionally, we examined the impact of $\tan\beta$ and the COM energy on the production cross section of $A$. We found that although the cross section is larger in the high $\tan \beta$ region, the enhancement due to the NLO is most significant in the low $\tan \beta$ region, which is a due of the contribution from the top loop in Type-II, and from both the top and the bottom loops in Type-X, which become significant in that region. We also found that although the cross section exhibits the typical $1/s$ phase space suppression, the NLO enhancement is nonetheless most significant in the high $\sqrt{s}$ region, which is a result of the derivative-type coupling of the EFT vertices that lead to energy enhancement that compensates for the phase space suppression.

Furthermore, we have compared our findings with the constraints from the LHC and from the decay $B \rightarrow X_{s}\gamma$ and found that a significant portion of the parameter space in Type-II, especially for $m_{A} \lesssim 1500$ GeV, is excluded. On the other hand, the constraints on Type-X remain weak. In general, we found that for the open region of the parameter space, cross sections up to $\sim 1$ fb for Type-II, and $\sim 5$ fb for Type-X are possible in some regions of the parameter space. A key lesson of this study is that muon colliders represent an excellent venue for searching for the the pseudoscalar $A$ in the 2HDM.

In this study, we confined our analysis to the Type-II and Type-X 2HDM, however, similar analysis can be easily extended to the other types of the 2HDM. In addition, it is interesting to investigate the possibility of distinguishing the 4 different 2HDMs at small $\tan\beta$. It was demonstrated in \cite{Han:2021udl} that muon colliders furnish an excellent probe to these different models for $\tan\beta \gtrsim 5$, and it would be interesting to investigate whether this capability can be extended to $\tan\beta \lesssim 5$. We postpone this to future work.

\begin{acknowledgments}
We thank Biplob Bhattacharjee for a useful discussion on the manuscript. 
Sagar Modak acknowledges the support of the Department of Science and Technology (DST), Government of India, through the INSPIRE fellowship, which has significantly contributed to the advancement of this research. The work of Samadrita Mukherjee is supported by IOE-IISc fellowship. SKV is supported by SERB, DST, Govt. of India Grants MTR/2022/000255, ``Theoretical aspects of some physics beyond standard models'', CRG/2021/007170 ``Tiny Effects from Heavy New Physics'' and IoE funds from IISC. 
\end{acknowledgments}

\bibliography{main}
\bibliographystyle{JHEP}
\end{document}